\documentclass[aps,amsmath,amssymb,preprint]{revtex4}
\usepackage{graphicx}
\usepackage{dcolumn}
\usepackage{bm}
\usepackage[utf8]{inputenc}
\usepackage[T1]{fontenc}
\usepackage{mathptmx}
\def\erfc{\mathop{\rm erfc}\nolimits}
\newtheorem{definition}{Definition}
\newtheorem{example}{Example}
\newtheorem{theorem}{Theorem}
\newtheorem{proposition}{Proposition}
\begin{document}
\title{Separatrices in the Hamilton-Jacobi Formalism of Inflaton Models}
\author{Gabriel \'Alvarez}
 \email{galvarez@ucm.es}

\author{Luis Mart\'{\i}nez Alonso}
\email{luism@ucm.es}
\affiliation{
Departamento de F\'{\i}sica Te\'orica, Facultad de Ciencias F\'{\i}sicas, Universidad Complutense de Madrid, 280240 Madrid, Spain
}

\author{Elena Medina}
\email{elena.medina@uca.es}
\affiliation{
Departamento de Matem\'aticas, Facultad de Ciencias, Universidad de C\'adiz, 11510 Puerto Real, C\'adiz, Spain
}

\author{Juan Luis V\'azquez}
\email{juanluis.vazquez@uam.es}
\affiliation{
Departamento de Matem\'aticas, Universidad Aut\'onoma de Madrid, 28049 Madrid, Spain
}
\date{\today}
\begin{abstract}
We consider separatrix solutions of the differential equations for inflaton models with a single scalar field
in a zero-curvature Friedmann-Lema\^{\i}tre-Robertson-Walker universe.
The existence and properties of separatrices are investigated in the framework of the Hamilton-Jacobi formalism,
where the main quantity is the Hubble parameter considered as a function of the inflaton field.
A wide class of inflaton models that have separatrix solutions (and include many of the most physically
relevant potentials) is introduced, and the properties of the corresponding separatrices are investigated,
in particular, asymptotic inflationary stages, leading approximations to the separatrices, and full
asymptotic expansions thereof. We also prove an optimal growth criterion for potentials that do not
have separatrices.
\end{abstract}
\maketitle
\section{Introduction\label{sec:intro}}
The theory of inflation  is a candidate to solve several long-standing problems regarding the physical
conditions in the early universe~\cite{ST80,GU81,LI82}. The present paper deals with  single-field inflationary
models defined by a potential function $V(\Phi)$ of the inflaton field  $\Phi$ in
a spatially flat Friedmann-Lema\^{\i}tre-Robertson-Walker  spacetime~\cite{MU05,BA09}.
The dynamical equation of these models  is the nonlinear ordinary differential equation
\begin{equation}
 	\label{eq:ids1}
	\ddot{\Phi}+3H\dot{\Phi}+V'(\Phi)=0,
\end{equation}
where $H$ is the Hubble parameter
\begin{equation}
	\label{eq:hs}
	H^2=\frac{1}{3\mathrm{M_{Pl}}^2}\left(\frac{1}{2} \dot{\Phi}^2 +V(\Phi)\right),
\end{equation}
$\mathrm{M_{Pl}}$ is the reduced Planck mass, dots denote derivatives with respect to the cosmic time $t$,
and primes denote derivatives of a function with respect to its argument.
In turn, the Hubble parameter $H$ is defined in terms of the scale factor $a$ by $H=\dot{a}/a$.
We consider models describing an expanding universe, for which $H$ is strictly positive at all times
and the inflationary stage is characterized by the condition $\ddot{a}>0$.

It is useful to  rescale variables
\begin{equation}\label{eq:ch}
	 	\Phi(t) = \sqrt{\frac{2}{3}}\,\mathrm{M_{Pl}}\,\varphi(t),\quad V(\Phi)=\frac{\mathrm{M_{Pl}}^2} {3}v(\varphi),\quad H(t) = \frac{1}{3} h(t),
\end{equation}
so that~(\ref{eq:ids1}) and~(\ref{eq:hs}) can be written as
\begin{equation}
	\label{eq:me}
	\ddot{\varphi} + h \dot{\varphi} + \frac{1}{2}v'= 0,
\end{equation}
and
\begin{equation}
	\label{eq:htau}
	h= (\dot{\varphi}^2+ v(\varphi))^{1/2},
\end{equation}
respectively. In terms of $h$ the inflation condition $\ddot{a}>0$ reads
\begin{equation}
	\label{ih}
	h<\sqrt{\frac{3v}{2}}.
\end{equation}

It is often claimed in the literature that many inflationary models exhibit ``attractor solutions'' of the differential equation~(\ref{eq:me})
to which solutions evolve for wide sets of initial conditions~\cite{BE85,LP94}. This terminology has been contested by
several authors~\cite{LP94,RC13},  since the notion of attractor used in those contexts does not correspond to the mathematical notion
of attractor of a flow in the theory of dynamical systems as defined, for example, in reference~\cite{HU82}, and in particular does not satisfy
Liouville's theorem on attractor behavior in Hamiltonian systems~\cite{RC13}. (On the other hand, the origin in the $(\varphi,\dot{\varphi})$
phase plane is an attractor in the mathematical sense, as is any point $(\varphi_0,0)$ where $\varphi_0$ is a point of minimum of the potential.)

The solutions often called attractors are in fact \emph{separatrices}~\cite{BE85}, and near these separatrices occur the crucial
inflationary stages of the solutions of~(\ref{eq:me}).

There is an extensive body of work on dynamical systems theory as applied to Cosmology and in particular to
Inflationary Cosmology.~\cite{BE85,CO98,FO98,UG13,FR14,RO14,TA14,PA15,GA15E,AU15,AU15D,AU17,BA18}
The standard approach to study (\ref{eq:ids1})--(\ref{eq:hs}) uses a Poincar\'e compactification of the $(\Phi, \dot{\Phi})$ phase plane
which resolves singular points at infinity and is specially suited to obtain global dynamical results and therefore all
possible asymptotic behaviors. Particularly relevant to our work are the studies in references~\cite{BE85,UG13,AU15},
which discuss orbits connecting critical points.

In this paper, however, we adopt the Hamilton-Jacobi formalism used by Salopek and Bond~\cite{SA90}
to study the evolution of long-wavelength metric fluctuations and to find (in parametric form) the general isotropic
solution of the exponential model known to drive power-law inflation, and by Liddle, Parsons and Barrow~\cite{LP94}
to generate explicit slow-roll expansions. More recently, the Hamilton-Jacobi formalism has been  also successfully applied
by Handley \emph{et. al.}\cite{HA14} to  study inflationary  solutions of general inflationary models emerging from
regions of kinetic dominance.

We stress that unlike the dynamical system approach, our straightforward implementation
of the Hamilton-Jacobi method is not global (in particular because of the lack of Poincar\'e compatification), but
on the other hand it permits using concepts and methods from the theory of ordinary differential equations such as super-solutions
and isoclines~\cite{BI89,TE12} to characterize separatrices and generate efficiently their asymptotic expansions.

Independently of the formalism used, the study of separatrices is linked to the analysis of singularities. It follows from equation~(\ref{eq:me}) that
\begin{equation}
	\label{hp}
	\dot{h} = -\dot{\varphi}^2,
\end{equation}
and therefore the Hubble parameter  $h$ is a positive monotonically decreasing function of $t$.
This property implies that  for smooth and positive potentials $v(\varphi)$,  the solutions $\varphi(t)$  of~(\ref{eq:me})
with arbitrary finite initial data  do not have singularities forward in the cosmic time $t$.  Furthermore,  it can be proved that under
mild conditions~\cite{ON71,ON78,MU82}, as $t\rightarrow\infty$ the solutions $\varphi(t)$ of~(\ref{eq:me})  tend to critical points of $v(\varphi)$.
However, due to~(\ref{hp}) the function $h(t)$ may increase without bound backwards in time,
so that $h(t)$ and $\varphi(t)$ may develop  singularities. The presence of singularities backwards in time can be also expected from
the following heuristic argument: If the  condition
\begin{equation}
	\label{kin}
	\dot{\varphi}^2 \gg v(\varphi),
\end{equation}
holds then we may neglect $v$ and $v'$   in the inflaton equations~(\ref{eq:me})--(\ref{eq:htau}) and approximate~(\ref{eq:me})  by
\begin{equation}
	\label{appr}
	\ddot{\varphi} +  |\dot{\varphi}  |\, \dot{\varphi} \sim 0.
\end{equation}
Hence we obtain two families of  approximate solutions
\begin{equation}
	\label{app-}
	\varphi(t)\sim -\log (t-t^*)+A,\quad t\rightarrow t^* \quad \mbox{for $\dot{\varphi}<0$ },
\end{equation}
and
\begin{equation}
	\label{app+}
	\varphi(t)\sim \log (t-t^*)+A,\quad t\rightarrow t^* \quad \mbox{for $\dot{\varphi}>0$ },
\end{equation}
 with $t>t^*$. These solutions depend on  two arbitrary parameters $t^*$ and $A$ which determine movable logarithmic singularities.
 The corresponding Hubble parameter~(\ref{eq:htau}) satisfies
\begin{equation}
	\label{app2}
	h(t)\sim \frac{1}{t-t^*} ,\quad  t\rightarrow t^*.
\end{equation}
For both approximate solutions~(\ref{app-})--(\ref{app+}), we have that $\dot{\varphi}^2 \sim \exp{(2 |\varphi|)}$, and in view of~(\ref{eq:htau}),
these singularities should not arise  for confining potentials that grow faster than $\exp{(2 |\varphi|)}$ as $|\varphi|\rightarrow \infty$~\cite{HA14}.
Condition~(\ref{kin}) defines what is called the  kinetic dominance regime of inflation~\cite{DE10, HA14, HA19}, and the separatrices,
if they exist, are boundaries of the regions in the $(\varphi,\dot{\varphi})$ phase space filled by the solutions with either asymptotic
behavior~(\ref{app-}) or~(\ref{app+}).

In the Hamilton-Jacobi formulation of inflationary models~\cite{SA90,LP94, BA09,LL09}
the basic dependent variable is the Hubble parameter $h=h(\varphi)$ as a function of the inflaton field $\varphi$.
Our main goal is to discuss the existence and characterization of separatrices from the large-$\varphi$ behavior of $h(\varphi)$,
which in turn will lead us to algorithms for the calculation of complete asymptotic expansions.

The paper is organized as follows. In Section~\ref{sec:v2} we present the equations of the Hamilton-Jacobi formalism
and its phase space $R$. We formulate sufficient conditions for the presence of solutions $\varphi(t)$ that
blow up at a finite time. We also show that when restricted to $R$, solutions can be classified into two types
which determine two non-overlapping regions filling the phase space $R$, and separatrices
will be defined as boundaries between these two regions. Section~\ref{sec:shj} shows that if there exists
a solution such that $h(\varphi)/\sqrt{v(\varphi)}$ is defined and bounded for large $\varphi$, then this solution is unique and
is a (the) separatrix. The actual question of existence of separatrices is discussed in Section~\ref{sec:exsep}, where a wide class of potentials
(which includes the standard potentials found in the literature) with separatrices is introduced. In addition, several examples of
inflationary models beyond this class both with and without separatrices are also exhibited.
Finally, Section~\ref{sec:asy} is devoted to the study of the asymptotics of separatrices.

Our main results are
\begin{enumerate}
	\item  We prove that the separatrix solutions $h_\mathrm{s}=h_\mathrm{s}(\varphi)$ separate  bounded from unbounded trajectories
		  in  the phase space $R$. In fact, and because of a symmetry discussed below, it is enough to focus on
		  unbounded trajectories in $R$ as $\varphi\to\infty$, which correspond to the singular solutions~(\ref{app-}).
	\item We determine a wide class $\mathcal{C}_{ \alpha}\, (0\leq\alpha<1)$ of potential functions which
	          determine inflationary models with separatrices. In particular, the even monomial potentials, the Higgs potential
	         and the Starobinsky potential are members of $\mathcal{C}_{ 0}$.
	         We also exhibit confining potentials without separatrices and
	         potentials outside the class $\mathcal{C}_{ \alpha}$ with a different type of separatrix solutions,
	         for which $h(\varphi)/\sqrt{v(\varphi)}$ is not bounded as $\varphi\to\infty$.
	\item We calculate the leading term  of the asymptotic expansion as $\varphi\rightarrow \infty$ of separatrices
	         for models of the class $\mathcal{C}_{ \alpha}$. This result is applied to prove that backwards inflation
	         only occurs in the separatrix solutions of these models for $\alpha<1/\sqrt{3}$.
	\item For asymptotically divergent potentials with $\alpha=0$ we obtain recursive relations that permit
	         to calculate explicitly as many terms as desired of the complete expansions of the separatrices
	         in terms of differential polynomials of the square root $\sqrt{v}$.
	         We also give asymptotic expansion for $\alpha$-attractor E-models and models with
	         hard potential walls.
	\item We argue heuristically that these asymptotic expansions are likely to be Borel summable
	         and discuss algorithms for an efficient approximate summation of these series.
	\item Finally, we analize the presence or absence of blow up in the inflaton fields $\varphi_\mathrm{s}(t)$
	         corresponding to the separatrix solutions of several basic models.
\end{enumerate}
\section{The Hamilton-Jacobi formalism\label{sec:v2}}
In this section we assume that the scaled potential $v=v(\varphi)$ is a smooth positive function.
\subsection{The Hamilton-Jacobi formalism and its phase space}
The main physical quantity in the so-called Hamilton-Jacobi formalism for inflationary models
is the scaled Hubble parameter $h(\varphi)$ considered as a function of the inflaton $\varphi$,
which satisfies
\begin{equation}
	\label{difin}
	(h')^2 = h^2 - v,
\end{equation}
and  gets its time dependence through the dependence of the inflaton $\varphi$ on the cosmic time $t$.
In essence and due to equation~(\ref{eq:htau}), this idea requires to consider $\dot{\varphi}$ as a function of $\varphi$,
and therefore to deal separately with each strictly monotonic part of $\varphi(t)$.

To formalize this idea, consider the lower and upper half-planes in the $(\varphi,\dot{\varphi})$ phase plane,
\begin{equation}
	\label{dpm}
	D_{-}=\{(\varphi,\dot{\varphi})\in \mathbb{R}^2 :  \dot{\varphi}<0\},
	\quad
	D_{+}=\{(\varphi,\dot{\varphi})\in \mathbb{R}^2 :  \dot{\varphi}>0\}.
\end{equation}
The map
\begin{equation}
	\label{ps}
	(\varphi,\dot{\varphi}) \mapsto (\varphi,h),
\end{equation}
defines two diffeomorphisms $T_{\pm} : D_{\pm} \rightarrow R$ from $D_{\pm}$ onto the open set
\begin{equation}
	\label{Rpm}
	R=\{(\varphi,h)\in \mathbb{R}^2 : \sqrt{v(\varphi)}<h<+\infty\},
\end{equation}
of the $(\varphi,h)$ plane. It follows from equation~(\ref{hp}) that
\begin{equation}
	\label{e}
	\dot{\varphi} = -h'(\varphi),
\end{equation}
and that the parts of each solution $\varphi(t)$ of equation~(\ref{eq:me}) in $D_{-}$ (strictly decreasing)
and in $D_{+}$ (strictly increasing) are described by the differential equations
\begin{equation}
	\label{hpm1}
	h'=\sqrt{h^2-v}, \quad (\varphi,h)\in R,
\end{equation}
and
\begin{equation}
	\label{hpm2}
	h'= -\sqrt{h^2-v}, \quad (\varphi,h)\in R,
\end{equation}
respectively. The two first-order non-linear ordinary differential equations~(\ref{hpm1})--(\ref{hpm2})
along with equation~(\ref{e}) are referred to as the Hamilton-Jacobi formalism for inflationary models~\cite{SA90,LP94, BA09,LL09}.

\begin{figure}	
		\includegraphics[width=8cm]{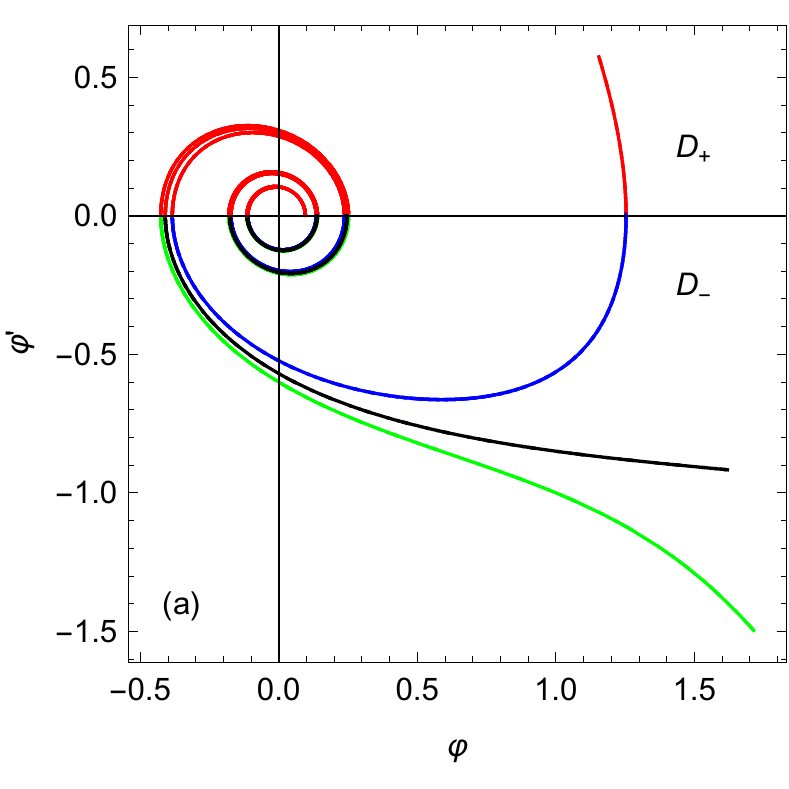}
		\includegraphics[width=8cm]{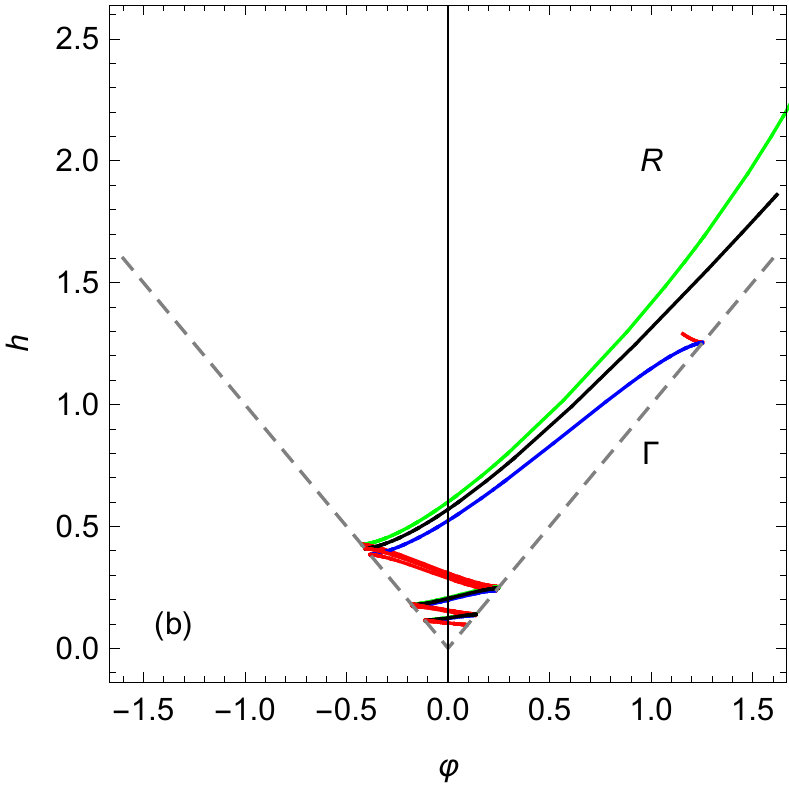}
	\caption{Three trajectories in the $(\varphi,\dot{\varphi})$ phase plane  of the quadratic model $v(\varphi)=\varphi^2$
	              exhibiting several monotonic pieces, and their corresponding arcs satisfying equations~(\ref{hpm1}) or~(\ref{hpm2})
	              in the phase space $R$ .\label{fig:fig1}}
\end{figure}

A main advantage of this Hamilton-Jacobi formalism is that if we ignore the time dependence given by
equation~(\ref{e}) and restrict our attention to the Hamilton-Jacobi phase plane $R$,
equations~(\ref{hpm1}) and~(\ref{hpm2}) are precisely the equations for the orbits of the dynamical system~(\ref{eq:me}).
Note that a given solution $\varphi(t)$ may consist of several (even infinitely many) strictly monotonic pieces,
each piece in $D_-$ and $D_+$ leading to a corresponding monotonic arc $h(\varphi)$ in the Hamilton-Jacobi phase
space $R$~(see Fig.~\ref{fig:fig1}) satisfying  equation~(\ref{hpm1}) and~(\ref{hpm2}), respectively.
We call the curve $h=\sqrt{v(\varphi)}$ the lower boundary $\Gamma$ of $R$, and note that the limiting value of $h'(\varphi)$
as $\varphi\to\Gamma$ is zero, i.e., if a solution $h(\varphi)$ reaches the lower boundary $\Gamma$,
it does so with an horizontal half-tangent line.

An initial value problem in the Hamilton-Jacobi formalism is determined by a point $(\varphi(0),h(\varphi(0)))$ in $R$
and by the sign of $\dot{\varphi}(0)$. For example, if  $\dot{\varphi}(0)$ is negative, its value
in terms of $\varphi(0)$ and $h(\varphi(0)))$ is given by equation~(\ref{hpm1}),
\begin{equation}
	\label{dop}
	\dot{\varphi}(0)=- h'(\varphi(0))=- \sqrt{h(\varphi(0))^2-v(\varphi(0))},
\end{equation}
and the solution $\varphi(t)$ is implicitly defined by
\begin{equation}
	\label{imp}
	t=-\int_{\varphi(0)}^{\varphi(t)}\frac{d \varphi}{h'(\varphi)}
\end{equation}
in the corresponding interval where $h'(\varphi)>0$.

The  Hamilton-Jacobi formalism is particularly convenient for studying the possible blow up of solutions $\varphi(t)$.
Indeed, the solution $\varphi(t)$ defined by~(\ref{imp}) blows up at a given
$t^*<0$ (i.e., $\varphi(t)\rightarrow +\infty$ as $t\rightarrow t^*$) if and only if the integral
\begin{equation}
	\label{impb}
	t^*=-\int_{\varphi(0)}^{\infty}\frac{d \varphi}{h'(\varphi)}
\end{equation}
is well-defined and convergent on that interval. In this case, the orbit $(\varphi,\dot{\varphi})$
tends to infinity in the fourth quadrant (e.g., the green orbit in Fig.~\ref{fig:fig1}).
Note that from equations~(\ref{app-}) and~(\ref{app2}) it follows that
\begin{equation}
	h(\varphi) \sim e^{\varphi- A},\quad\varphi\to\infty.
\end{equation}
From~(\ref{impb}) it is also clear that a necessary condition for a solution $h=h(\varphi)$ of~(\ref{hpm1})
to determine a blow-up solution $\varphi=\varphi(t)$   is that
\begin{equation}\label{lim}
\lim_{\varphi\rightarrow \infty}\Big(h(\varphi)^2- v(\varphi)\Big)=\infty.
\end{equation}

Similar statements can be made for solutions with $\dot{\varphi}(0)>0$ and equation~(\ref{hpm2}).  In particular,
orbits $(\varphi,\dot{\varphi})$ corresponding to blow-up solutions tend to infinity in the second quadrant.
Since~(\ref{hpm2}) reduces to~(\ref{hpm1}) under the change of variable
\begin{equation}
 	\label{cha}
 	\widetilde{h}(\varphi)=h(-\varphi), \quad \widetilde{v}(\varphi)=v(-\varphi),
 \end{equation}
the analysis of~(\ref{hpm1}) can be also applied to  the strictly monotonic parts of  the solutions  $\varphi(t)$ of the inflaton
equation~(\ref{eq:me}) with trajectories in $D_+$. Therefore, hereafter we will focus our attention on the differential
equation~(\ref{hpm1}).  Thus we will only deal with monotonic parts of  solutions  $\varphi(t)$  lying in the fourth quadrant
of the plane $(\varphi,\dot{\varphi})$.
\subsection{The two types of possible orbits in the Hamilton-Jacobi formalism}
Our main purpose is to characterize separatrices that extend to infinity in $R$. Therefore, it is enough to consider
regions of the phase space $R$ to the right of some appropriate (potential-dependent) value $\varphi_0$,
i.e., regions of the form
\begin{equation}
	\label{dom2n}
	R_0=\{(\varphi,h)\in \mathbb{R}^2\, :\, \varphi_0 \leq \varphi <+\infty,\, \sqrt{v(\varphi)}<h<+\infty\},
\end{equation}
where we assume that  the scaled potential $v$ and its first derivative $v'$ are smooth
and strictly positive  for $\varphi\geq \varphi_0$ . Note that this stipulation extends the validity of our analysis to potentials not necessarily
monotonic on the whole real line.

The next proposition states that the solutions $h(\varphi)$ of equations~(\ref{hpm1}) in $R_0$ do not blow
up at finite values of $\varphi$.
\begin{proposition}
The solution of~(\ref{hpm1}) with initial value $ h(\varphi_0)=h_0$ satisfies
\begin{equation}
	\label{concc}
	h(\varphi)<h_0 e^{\varphi-\varphi_0},\quad \forall \varphi>\varphi_0.
\end{equation}
\end{proposition}

\noindent
\emph{Proof:}
Taking into account that $\sqrt{h^2-v}< h$, the  solutions  $h_\mathrm{sup}(\varphi)=h_0 e^{\varphi-\varphi_0}$
of the  differential equation
\begin{equation}
	\label{sbs}
	h_\mathrm{sup}'=h_\mathrm{sup},
\end{equation}
are super solutions~\cite{BI89,TE12} of equation~(\ref{hpm1}). Hence given solutions $h$ of~(\ref{hpm1})
and $h_\mathrm{sup}$ of~(\ref{sbs}) with the same  initial data $(\varphi_0,h_0)\in R_0 $, then $h(\varphi)<h_\mathrm{sup}(\varphi)$
for $ \varphi>\varphi_0$, i.e., any solution $h(\varphi)$ is bounded by a suitable exponential and
therefore cannot blow up at a finite value of $\varphi$. $\square$

The next proposition is a straightforward consequence of Proposition~1, and shows that there are only two possible behaviors
for the orbits, which we call type-A and type-B solutions of equation~(\ref{hpm1}).

\begin{proposition} Given a  solution $h=h(\varphi)$  of~(\ref{hpm1}) with initial value $h(\varphi_0)=h_0$, then  either it leaves $R_0$
by reaching the lower boundary $\Gamma$ with a horizontal half-tangent line (type-A solution), or it exists and remains in $R_0$
for all $\varphi> \varphi_0$ (type-B solution).
\end{proposition}

Some comments are in order.
\begin{enumerate}
	\item Since $\sqrt{h^2-v}$ is smooth and positive on $R_0$, the solution $h(\varphi)$ obtained by integrating
	equation~(\ref{hpm1}) backwards from any point $(\varphi_1, \sqrt{v(\varphi_1)})$ of $\Gamma$
	 can be continued all the way back to the vertical line $\varphi=\varphi_0$ of $R_0$.
	Since this solution has zero slope at $\varphi_1$ (because the differential equation~(\ref{hpm1})
	 implies $h'|_{\Gamma}=0$), by monotonicity it cannot end at another point
	 $(\varphi_2, \sqrt{v(\varphi_2)})$ of $\Gamma$ with $\varphi_2<\varphi_1$.
	\item The set of  solutions that end at the lower boundary $\Gamma$ (type-A solutions)
	is always nonempty and  fills a subregion of $R_0$ that may or may not be the
        whole phase space $R_0$. Again by monotonicity, it follows that there is an $r$ with $\sqrt{v(\varphi_0)}<r\leq \infty$
        such that the solutions starting at $h(\varphi_0)<r$ are type-A solutions. If $r\neq \infty$ the solutions with $h(\varphi_0)>r$ (type-B solutions)
        are global, i.e., they are defined for all $\varphi\geq \varphi_0$. As we will prove below,  for a wide class of models
        (the class $\mathcal{C}_{ \alpha}$ introduced below) these type-B solutions grow exponentially as $\varphi\rightarrow \infty$.
        Furthermore, the solution corresponding to $h(\varphi_0)=r$, which separates the two types of solutions,
        will be shown to be globally defined, although its rate of growth as $\varphi \rightarrow \infty$ depends on the potential of the model.
        This solution, when it exists, is what we call the separatrix of the model.
\end{enumerate}

\begin{figure}
		\includegraphics[width=8cm]{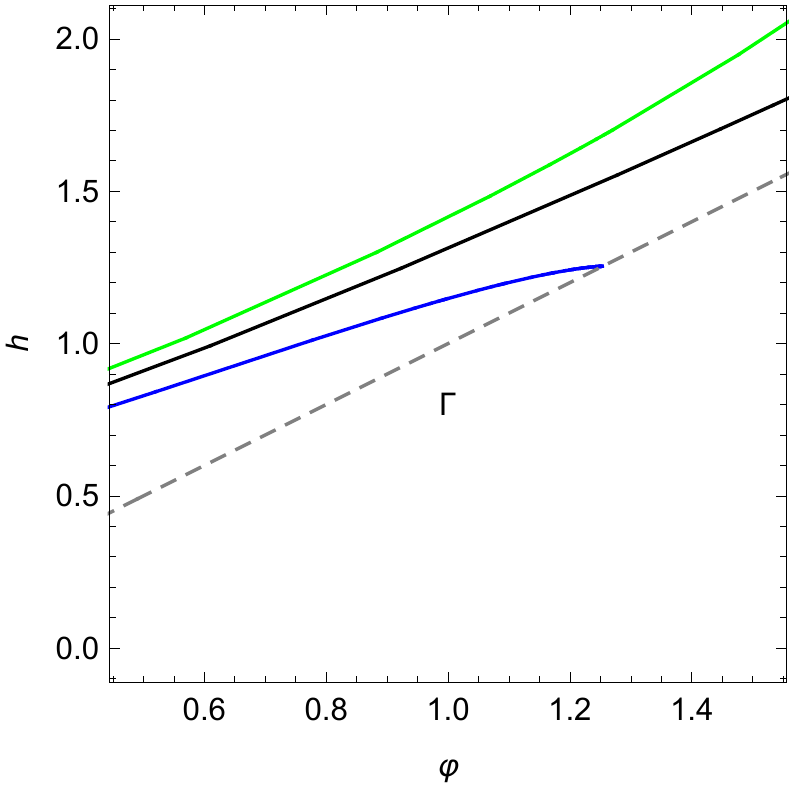}	
	\caption{Three solutions $h(\varphi)$ of  the equation~(\ref{hpm1}) in the phase space $R_0$ for the
	              quadratic model $v(\varphi)=\varphi^2$: the blue line is a type-A solution,
	              the green line is a type-B solution,  and the black  line is the separatrix (itself a type-B solution).
	              The grey line is the lower boundary $\Gamma$ of $R_0$.\label{fig:fig2}}
\end{figure}

\subsection{Separatrices}
We are now able to formulate a precise definition of a separatrix of equation~(\ref{hpm1}):
\begin{definition}
If equation~(\ref{hpm1}) has both type-A and type-B solutions, then $r<\infty$ and the solution corresponding
to the initial condition $h(\varphi_0)=r$ is called a separatrix.
\end{definition}
In  Section IV we will see that many potentials have separatrices for which $h(\varphi)/\sqrt{v(\varphi)}$
is bounded. It will be shown that  separatrices of this kind have special properties.
\section{Properties of separatrices in the Hamilton-Jacobi formalism\label{sec:shj}}
\begin{figure}
		\includegraphics[width=8cm]{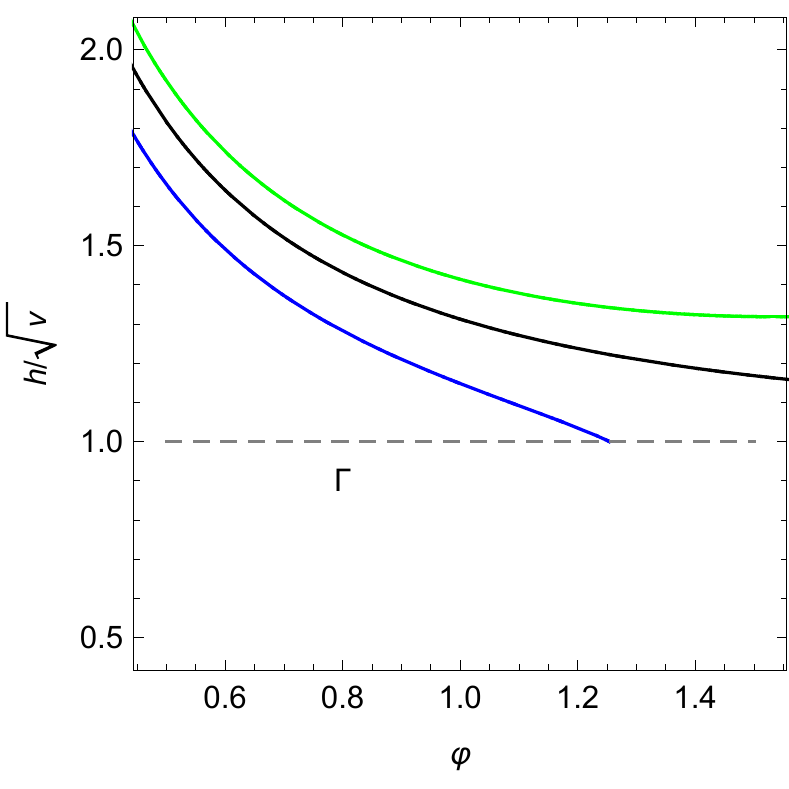}	
	\caption{Solutions $\mathfrak{h}(\varphi)$ of equation~(\ref{ymas}) in the phase space $\mathfrak{R}_0$
	              corresponding  to the curves $h(\varphi)$ of Fig.~2.\label{fig:fig3}}
\end{figure}
To gain a better understanding of the solutions of equation~(\ref{hpm1}) and of the existence of separatrices
it is useful to introduce the modified Hubble parameter
\begin{equation}
	\label{change}
	\mathfrak{h}(\varphi)\equiv \frac{h(\varphi)}{\sqrt{v(\varphi)}}.
\end{equation}
In terms of $\mathfrak{h}=\mathfrak{h}(\varphi)$ the orbit equation~(\ref{hpm1}) reads
\begin{equation}
	\label{ymas}
	\mathfrak{h}'=\sqrt{\mathfrak{h}^2-1}-\mathfrak{v}\,\mathfrak{h},
\end{equation}
where
\begin{equation}
	\label{iu}
 	\mathfrak{v}\equiv (\log \sqrt{v})'.
\end{equation}

Note that the function $\mathfrak{h}$ is related to the Hubble normalized potential $V/3H^2$: in fact
\begin{equation}
	\label{hn}
	\frac{1}{\mathfrak{h}^2} = \frac{v}{h^2} = \frac{V}{3 H^2},
\end{equation}
where we have set $\mathrm{M_{Pl}}=1$. Similarly, for positive, strictly increasing potentials $V(\Phi)$,
the function $\mathfrak{v}(\varphi)$ is proportional to the function $\lambda(\Phi)$ defined by
\begin{equation}
	\label{eq:lambda}
	\lambda(\Phi) = \frac{V'(\Phi)}{V(\Phi)}.
\end{equation}
Using the scaled variables~(\ref{eq:ch}) it follows that
\begin{equation}
	\label{lambdaiu}
	\lambda(\Phi) = \sqrt{\frac{3}{2}}\frac{v'(\varphi)}{v(\varphi)}=\sqrt{6}\mathfrak{v}(\varphi),
\end{equation}
again with $\mathrm{M_{Pl}}=1$. (Note that in the application of dynamical systems theory to
$\Lambda$CDM cosmology it is customary to use strictly decreasing potentials $V(\Phi)$
and to define $\lambda(\Phi)$ with an additional minus sign~\cite{AU15D}.)

The phase space of~(\ref{ymas}) is given by $\mathfrak{R}_0$
\begin{equation}
	\label{dom2ny}
	\mathfrak{R}_0=\{(\varphi,\mathfrak{h})\in \mathbb{R}^2\, :\, \varphi_0 \leq \varphi <+\infty,\, 1\leq \mathfrak{h}<+\infty\}.
\end{equation}

Incidentally, the function
\begin{equation}
	\label{ze}
	y\equiv \log \mathfrak{h}(\varphi),
\end{equation}
verifies the so-called \emph{master equation} introduced by Handley et al.~in~\cite{HA14}
\begin{equation}
	\label{mas}
	y'=\sqrt{1-e^{-2y}}-\mathfrak{v}.
\end{equation}

The equivalent of Proposition~2 for the modified Hubble parameter is:
\begin{proposition} A solution $\mathfrak{h}=\mathfrak{h}(\varphi)$  of~(\ref{ymas}) with initial value
$\mathfrak{h}(\varphi_0)=\mathfrak{h}_0$ either leaves $\mathfrak{R}_0$ by reaching the finite boundary
$\mathfrak{h}=1$ with negative slope $-\mathfrak{v}$ (type-A solution), or it exists and remains in $\mathfrak{R}_0$ for all
$\varphi> \varphi_0$ (type-B solution).
\end{proposition}

We illustrate this modified phase space $\mathfrak{R}_0$ and Proposition~3 in Fig.~\ref{fig:fig3},
where we plot the solutions $\mathfrak{h}(\varphi)$ of equation~(\ref{ymas}) with $\mathfrak{v}(\varphi)=1/\varphi$
corresponding to the curves $h(\varphi)$ of Fig.~2.

Equation~(\ref{ymas}) can be rewritten in the integral form
\begin{equation}
	\label{inte}
	 \mathfrak{h}(\varphi)=\frac{\sqrt{v(\varphi_0)}}{\sqrt{v(\varphi)}}\mathfrak{h}(\varphi_0)\exp{\int_{\varphi_0}^{\varphi}\sqrt{1-\frac{1}{\mathfrak{h}(x)^2}}\, d x}.
\end{equation}
Thus, given two solutions $\mathfrak{h}_i(\varphi)\, (i=1,2)$ of~(\ref{ymas})  we have
\begin{equation}
	\label{y2}
	\frac{\mathfrak{h}_2(\varphi)}{\mathfrak{h}_1(\varphi)}=\frac{\mathfrak{h}_2(\varphi_0)}{\mathfrak{h}_1(\varphi_0)}\,
	\exp{\int_{\varphi_0}^{\varphi}\left(\sqrt{1-\frac{1}{\mathfrak{h}_2(x)^2}}-\sqrt{1-\frac{1}{\mathfrak{h}_1(x)^2}}\right)\, d x}.
\end{equation}
Note that the potential $v$ does not appear in the identity~(\ref{y2}). This property is very
convenient to analyze the behavior of the solutions of~(\ref{hpm1}).
\begin{proposition} Given two solutions $h_i(\varphi)\, (i=1,2)$ of~(\ref{hpm1}) such that $h_2(\varphi_0)>h_1(\varphi_0)$, then the corresponding functions
$\mathfrak{h}_i(\varphi)\, (i=1,2)$ satisfy
\begin{equation}
	\label{p1}
	\mathfrak{h}_2(\varphi)>\mathfrak{h}_1(\varphi),\quad \mbox{for all $\varphi\geq \varphi_0$},
\end{equation}
and
\begin{equation}
	\label{p2}
	\frac{\mathfrak{h}_2(\varphi)}{\mathfrak{h}_1(\varphi)}>\frac{\mathfrak{h}_2(\varphi')}{\mathfrak{h}_1(\varphi')},\quad \mbox{for all $\varphi> \varphi'>\varphi_0$}.
\end{equation}
\end{proposition}

\noindent
\emph{Proof:} The first statement~(\ref{p1}) is a consequence of the fact that the functions
$\mathfrak{h}_i(\varphi)\, (i=1,2)$ are solutions of the ordinary differential equation~(\ref{ymas})
with initial conditions satisfying $\mathfrak{h}_2(\varphi_0)>\mathfrak{h}_1(\varphi_0)$.
The second statement~(\ref{p2}) follows at once from~(\ref{p1}) and the identity~(\ref{y2}). $\Box$

The next  Theorem is  a reformulation of several results proved by Handley et al.~\cite{HA14}
\begin{theorem} Let $\mathfrak{h}_\mathrm{s}=\mathfrak{h}_\mathrm{s}(\varphi)$ be a  solution of~(\ref{ymas}) defined
and bounded for all $\varphi\geq \varphi_0$ in $\mathfrak{R}_0$. Then:
\begin{enumerate}
	\item   $\mathfrak{h}_\mathrm{s}$ is the only  solution of~(\ref{ymas}) defined and bounded for all $\varphi\geq \varphi_0$.
	\item If a solution $\mathfrak{h}=\mathfrak{h}(\varphi)$ of~(\ref{ymas}) is such that  $\mathfrak{h}(\varphi_0)>\mathfrak{h}_\mathrm{s}(\varphi_0)$,
	then $\mathfrak{h}$ is a type-B solution and grows exponentially  as $\varphi \rightarrow\infty$.
	\item If a  solution $\mathfrak{h}=\mathfrak{h}(\varphi)$ of~(\ref{ymas}) is such that  $\mathfrak{h}(\varphi_0)<\mathfrak{h}_\mathrm{s}(\varphi_0)$,
	then $\mathfrak{h}$ is a type-A solution.
\end{enumerate}
Therefore $\mathfrak{h}_\mathrm{s}(\varphi)$ is the separatrix.
\end{theorem}

\noindent
\emph{Proof:} Let $\mathfrak{h}_\mathrm{s}$ be a solution of~(\ref{ymas}) defined and bounded  for all $\varphi\geq \varphi_0$,
and let $\mathfrak{h}$ be a solution of~(\ref{ymas}) such that  $\mathfrak{h}(\varphi_0)>\mathfrak{h}_\mathrm{s}(\varphi_0)$,
then according to~(\ref{p2}) we have that
\begin{equation}
	\label{p4}
	\frac{\mathfrak{h}(\varphi)}{\mathfrak{h}_\mathrm{s}(\varphi)}
	>
	\Delta \equiv \frac{\mathfrak{h}(\varphi_0)}{\mathfrak{h}_\mathrm{s}(\varphi_0)}
	>1,\quad \mbox{for all $\varphi>\varphi_0$}.
\end{equation}
Consequently
\begin{equation}
	\label{p5}
	\sqrt{1-\frac{1}{\mathfrak{h}(x)^2}}-\sqrt{1-\frac{1}{\mathfrak{h}_\mathrm{s}(x)^2}}
	>
	\sqrt{1-\frac{1}{\Delta^2 \mathfrak{h}_\mathrm{s}(x)^2}}-\sqrt{1-\frac{1}{\mathfrak{h}_\mathrm{s}(x)^2}}.
\end{equation}
Furthermore,  the following function of $\mathfrak{h}_\mathrm{s}$
\begin{equation}
	\label{p6}
	f(\mathfrak{h}_\mathrm{s})\equiv \sqrt{1-\frac{1}{\Delta^2 \mathfrak{h}_\mathrm{s}^2}}-\sqrt{1-\frac{1}{\mathfrak{h}_\mathrm{s}^2}},
\end{equation}
is positive and  decreasing, so that $f(\mathfrak{h}_\mathrm{s})\geq C$, where $C=f(\mathfrak{h}_*)$, with
$\mathfrak{h}_*$ being the supremum of $\mathfrak{h}_\mathrm{s}(\varphi)$ on the interval $\varphi \geq \varphi_0$.
Therefore, from~(\ref{y2}) and~(\ref{p5}) we obtain that
\begin{equation}
	\label{p7}
	\frac{\mathfrak{h}(\varphi)}{\mathfrak{h}_\mathrm{s}(\varphi)}
	>
	\Delta\,\exp{\left(C(\varphi-\varphi_0)\right)},\quad \mbox{for all $\varphi\geq \varphi_0$},
\end{equation}
so that $\mathfrak{h}(\varphi)$ grows exponentially as $\varphi \rightarrow\infty$.
This proves the statements \emph{1} and \emph{2}. Regarding statement~\emph{3}, it is clear that given
a solution $\mathfrak{h}=\mathfrak{h}(\varphi)$ of~(\ref{ymas})  such that  $\mathfrak{h}(\varphi_0)<\mathfrak{h}_\mathrm{s}(\varphi_0)$,
then  it would be a bounded solution defined for all $\varphi\geq \varphi_0$ unless it leaves $\mathfrak{R}_0$  by crossing the lower boundary
$ \mathfrak{h}=1$ of $\mathfrak{R}_0$ at a finite value of $\varphi$. $\Box$

From statement~\emph{2} of Theorem~1 it follows that   type B solutions $h=h(\varphi)$ of~(\ref{hpm1}) determine blow-up solutions $\varphi=\varphi(t)$
of the inflaton equation~(\ref{eq:me}).
\section{Existence of separatrix solutions\label{sec:exsep}}
\subsection{A class $\mathcal{C}_{ \alpha}$ of potentials with separatrix solutions}
\begin{definition} Given  $0\leq \alpha <1$ we define $\mathcal{C}_{ \alpha}$ as the set of all the $C^{\infty}$ real functions $v=v(\varphi)$
such that
\begin{enumerate}
	\item Both $v=v(\varphi)$  an its derivative $v'=v'(\varphi)$ are strictly   positive for all $\varphi$ larger than some $\varphi_0$.
	\item The function $\mathfrak{v}$ defined in equation~(\ref{iu}) satisfies
	\begin{equation}
		\label{asv}
	\lim_{\varphi\rightarrow \infty}\mathfrak{v}(\varphi)= \alpha.
	\end{equation}
\end{enumerate}
For convenience we will  take $\varphi_0$  such that $\mathfrak{v}(\varphi)<1$ for all $\varphi>\varphi_0$.
\end{definition}

Note that using equation~(\ref{lambdaiu}) for potentials of the class $\mathcal{C}_{ \alpha}$,
condition~(\ref{asv}) is equivalent to
\begin{equation}
	\label{eq:lambdaiua}
	\lim_{\Phi\to\infty} \lambda(\Phi) = \sqrt{6}\alpha.
\end{equation}
Note also, for later applications, that the $0$-isocline of equation~(\ref{ymas}) in $\mathfrak{R}_0$
is generally given by
\begin{equation}
	\label{isoc}
 \mathfrak{h}_\mathrm{iso}(\varphi)=\frac{1}{\sqrt{1-\mathfrak{v}(\varphi)^2}},
\end{equation}
which for potentials $v\in \mathcal{C}_{ \alpha}$ has the finite limit
\begin{equation}
	\label{isop}
	\lim_{\varphi\rightarrow \infty}\mathfrak{h}_\mathrm{iso}(\varphi)=\frac{1}{\sqrt{ 1-\alpha^2}}.
\end{equation}

In the following example we present a a particular family of exponential potentials belonging to the class
$\mathcal{C}_{ \alpha}$ which will be used in the proof of Theorem~2.

\begin{example} The family of exponential  potential functions~\cite{HA87,SA90,CO98}
\begin{equation}
	\label{exp2}
 	v_{\alpha}(\varphi)=(1-\alpha^2) \,e^{2 \alpha\varphi}, \quad 0< \alpha<1,
\end{equation}
belong to the class $ \mathcal{C}_{ \alpha}$. They determine a constant function $\mathfrak{v}(\varphi)$
\begin{equation}
	\label{asve}
	\mathfrak{v}(\varphi)=\alpha.
\end{equation}
Their corresponding Hamilton-Jacobi equations~(\ref{hpm1})  have an explicit solution given by
\begin{equation}
	\label{cej2}
	h_\mathrm{s}(\varphi)=e^{\alpha \varphi},
\end{equation}
which according to Theorem~1 is a separatrix. Furthermore, in this case the modified Hubble parameter
$\mathfrak{h}_\mathrm{s}(\varphi)$~(\ref{change}) coincides with the $0$-isocline solution~(\ref{isoc})
and it is given by the constant value
\begin{equation}
	\label{const}
	\mathfrak{h}_\mathrm{s}(\varphi)=\mathfrak{h}_\mathrm{iso}(\varphi)=\frac{1}{\sqrt{ 1-\alpha^2}}.
\end{equation}
\end{example}

\begin{theorem}
	If $v\in \mathcal{C}_{ \alpha}$ then the differential equation~(\ref{hpm1}) has a separatrix solution $h_\mathrm{s}(\varphi)$.
\end{theorem}
\noindent
\emph{Proof:}
Let us consider the differential equation~(\ref{ymas}) for the modified Hubble parameter $\mathfrak{h}$ determined
by $\mathfrak{v}=(\log \sqrt{v})'$. From  Proposition~3  we have that the phase space  of~(\ref{ymas})
\begin{equation}
	\label{dom2ny2}
	\mathfrak{R}_0=\{(\varphi,\mathfrak{h})\in \mathbb{R}^2\, :\, \varphi_0 \leq \varphi <+\infty,\, 1\leq \mathfrak{h}<+\infty\},
\end{equation}
contains two possible types of solutions,  those leaving   $\mathfrak{R}_0$ by crossing the lower finite boundary $\mathfrak{h}=1$
(type-A solutions) and those staying  in  $\mathfrak{R}_0$ for all $\varphi> \varphi_0$ (type-B solutions).
Notice that $\mathfrak{h}'(\varphi)=-\mathfrak{v}(\varphi)<0$ at any point $(\varphi,1)$ in the lower boundary.

Let us denote by   $I_A$ and $I_B$   the subsets of real numbers $\mathfrak{h}_0=\mathfrak{h}(\varphi_0)>1$
corresponding to the initial data of the  solutions $\mathfrak{h}=\mathfrak{h}(\varphi)$ of~(\ref{ymas}) of type $A$ and type $B$,
respectively.  From Proposition~3 we have that
\begin{equation}
	\label{sss}
	I_A\cap I_B=\emptyset,\quad I_A\cup I_B=(1,+\infty).
\end{equation}
The subset $I_A$ is  nonempty (see comment i) after Proposition~2). To prove that the subset $I_B$ is also nonempty
we use the model associated with a potential $v_{\alpha_0}$ of the form~(\ref{exp2}) with $\alpha<\alpha_0<1$.
Then, since $v\in \mathcal{C}_{ \alpha}$ and due to~(\ref{asv}) we can always choose a $\varphi_0$ such that
\begin{equation}
	\label{un1}
	\mathfrak{v}(\varphi) <\alpha_0, \quad \mbox{ for all $\varphi>\varphi_0.$}
\end{equation}
Hence,  it follows that
\begin{equation}
	\label{un2}
	\sqrt{\mathfrak{h}^2-1}-\mathfrak{v}\,\mathfrak{h}>\sqrt{\mathfrak{h}^2-1}-\alpha_0\,\mathfrak{h},
	\quad \mbox{ for all $(\varphi, \mathfrak{h})\in \mathfrak{R}_0$}.
\end{equation}
This means that the solutions of the differential equation~(\ref{ymas}) corresponding to $\mathfrak{v}$
are super solutions of the differential equation~(\ref{ymas}) corresponding to $v_{\alpha_0}$.

We know that the constant line
\begin{equation}
	\label{un3}
	\mathfrak{h}_\mathrm{iso}(\varphi) = \frac{1}{\sqrt{1-\alpha_0^2}}
\end{equation}
is a $0$-isocline of the differential equation~(\ref{ymas}) corresponding to $v_{\alpha_0}$, and because of equation~(\ref{un2}),
the solutions of the differential equation~(\ref{ymas}) corresponding to $v$ which cross this line will do it with a positive value of $\mathfrak{h}'$.
Consequently they cannot come back below the line~(\ref{un3}), they are type-B solutions, and $I_B$ is nonempty.

\begin{figure}
		\includegraphics[width=8cm]{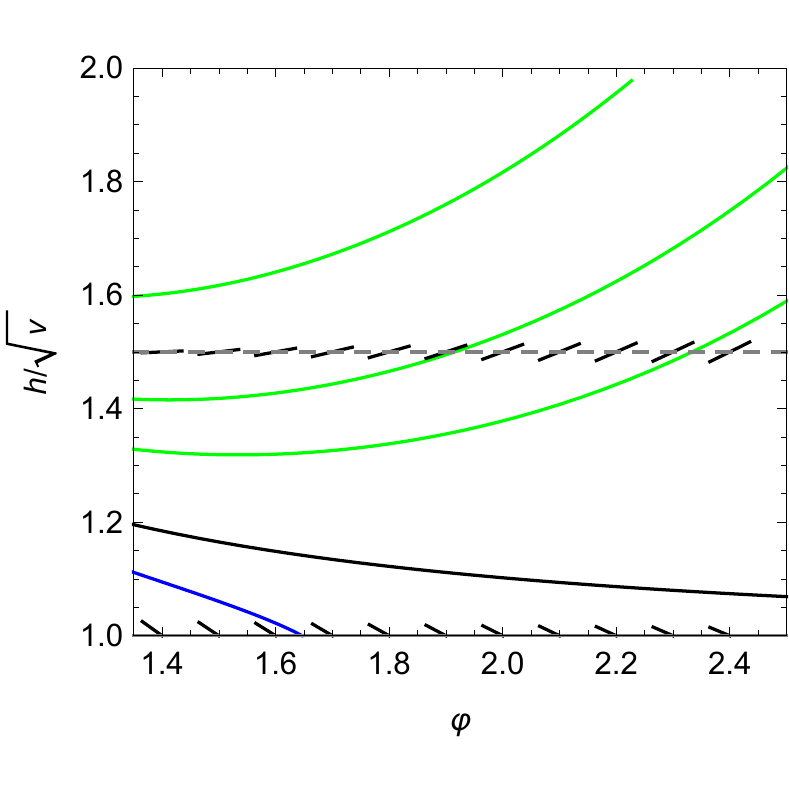}	
	\caption{Illustration of the argument leading to Theorem~2 for the case of the quadratic potential
	              $v(\varphi)=\varphi^2$ and $\alpha_0=\sqrt{5}/3$. The horizontal dashed line is the 0-isocline
	              $\mathfrak{h}_\mathrm{iso}(\varphi) = 3/2$, the slopes on top of this line correspond
	              to equation~(\ref{ymas}) with $\mathfrak{v}(\varphi) = 1/\varphi$ (as do the half-slopes at $\mathfrak{h}=1$),
	              and $\varphi_0$ has been taken slightly greater than $1/\alpha_0$.
	              The region between the horizontal lines $\mathfrak{h}=1$ and
	              $\mathfrak{h}_\mathrm{iso}(\varphi) = 3/2$ is a backwards-invariant region for the
	              flow generated by equation~(\ref{ymas}).
	              The separatrix is the black, continuous curve.\label{fig:fig4}}
\end{figure}

If we denote by $r$ the real number defining both the  supremun of the set $I_A$ and the infimun of the set $I_B$, then the solution
$\mathfrak{h}_r(\varphi)$ of~(\ref{ymas}) such that  $\mathfrak{h}_r(\varphi_0)=r$ is of type $B$. Otherwise, it would hit a point $\varphi_r$
of the boundary $ \mathfrak{h}=1$ and the backwards solution $\widetilde{ \mathfrak{h}} (\varphi)$ corresponding to another point
$\widetilde{\varphi}>\varphi_r$ of the boundary $ \mathfrak{h}=1$ would verify $ \widetilde{ \mathfrak{h}} (\varphi_0)>r$, which is a contradiction.
Furthermore, it can not cross the line~(\ref{un3}). Indeed, if it hits that line at a point  $(\widetilde{\varphi}_0, (1-\alpha_0^2)^{-1/2})$,
then the backwards solution $\widetilde{ \mathfrak{h}} (\varphi)$ corresponding to another point of the line~(\ref{un3}) to the right
of $(\widetilde{\varphi}_0, (1-\alpha_0^2)^{-1/2})$ will verify $ \widetilde{ \mathfrak{h}} (\varphi_0)<r$, which is a contradiction.
Finally, it is clear that  $\mathfrak{h}_r$ is a bounded solution as it is defined for all $\varphi\geq \varphi_0$ and remains inside
the region bounded  by the lines $ \mathfrak{h}=1$ and $ \mathfrak{h}=(1-\alpha_0^2)^{-1/2}$.
Therefore, $\mathfrak{h}_r(\varphi)$ is a separatrix solution. $\Box$

In Fig.~\ref{fig:fig4} we illustrate this argument for the quadratic potential $v(\varphi)=\varphi^2$, where we have taken
$\alpha_0=\sqrt{5}/3$, so that the horizontal dashed line corresponds to the 0-isocline $\mathfrak{h}_\mathrm{iso}(\varphi) = 3/2$,
while the slopes plotted on top of this line correspond to equation~(\ref{ymas}) with $\mathfrak{v}(\varphi) = 1/\varphi$ and $\varphi_0$
has been taken slightly greater than $1/\alpha_0$.

\begin{example}
The model with the Higgs potential~\cite{MA14}
\begin{equation}
	\label{exp1}
	v(\varphi)=(\varphi^2-1)^2,
\end{equation}
belongs to the extended class of potentials introduced in section II.B for $\varphi_0\geq1$, and  has an explicit  separatrix solution of~(\ref{hpm1})
given by
\begin{equation}
	\label{cej}
	h_\mathrm{s}(\varphi)=\varphi^2+1.
\end{equation}
Notice that  $\mathfrak{h}_\mathrm{s}(\varphi) = (\varphi^2+1)/(\varphi^2-1) \to 1\mbox{ as }\varphi\to\infty$.
\end{example}

\subsection{A class of potentials without a separatrix}
The next Proposition establishes an exponential lower bound  for models $v\not \in \mathcal{C}_{ \alpha}$ without  separatrix solutions.

\begin{proposition} If the potential $v(\varphi)$ grows  faster than $\exp{(2\varphi)}$  as $\varphi\rightarrow \infty$,
then all the solutions of equation~(\ref{hpm1}) are type-A solutions and there is no separatrix.
\end{proposition}

\noindent
\emph{Proof:}
As a consequence of Proposition 1 the functions  $h_\mathrm{sup}(\varphi)=h_0 e^{\varphi-\varphi_0}$  are super solutions
of~(\ref{hpm1}). Hence, under our assumption on  $v(\varphi)$,  these super solutions cross the boundary $ h(\varphi)=\sqrt{v(\varphi)}$
at finite values of $\varphi$, and so do the solutions of~(\ref{hpm1}),  which lie between  $\sqrt{v(\varphi)}$ and $h_\mathrm{sup}(\varphi)$ $\Box$

The sharp character of this bound is shown by the next example.
\begin{example}
Equation~(\ref{ymas}) for the potential function
\begin{equation}
	\label{exp20}
 	v(\varphi)=e^{2 \varphi}
\end{equation}
reads
\begin{equation}
	\label{mas1}
	\mathfrak{h}' = \sqrt{\mathfrak{h}^2-1}-\mathfrak{h}.
\end{equation}
The right-hand side of~(\ref{mas1}) is upper bounded by $-1/(2\mathfrak{h})$, so that the solutions
\begin{equation}\label{hss}
\mathfrak{h}_\mathrm{sup}(\varphi)=\sqrt{\mathfrak{h}_0^2+\varphi_0-\varphi},
\end{equation}
 of the  differential equation
\begin{equation}\label{sbs2}
\mathfrak{h}_\mathrm{sup}'=-\frac{1}{2 \mathfrak{h}_\mathrm{sup}},
\end{equation}
are super solutions  of~(\ref{mas1}). Hence if $\mathfrak{h}$ and $\mathfrak{h}_\mathrm{sup}$ are solutions of~(\ref{mas1}) and~(\ref{sbs2})
respectively,  with the same  initial data $(\varphi_0,\mathfrak{h}_0)\in \mathfrak{R}_0 $, then
$\mathfrak{h}(\varphi)<\mathfrak{h}_\mathrm{sup}(\varphi)$ for $ \varphi>\varphi_0$.
Thus, all the solutions of~(\ref{mas1}) leave $\mathfrak{R}_0$ at a finite value
of $\varphi$. Therefore, all the solutions are of type-A and, consequently, there is no separatrix solution for the model~(\ref{exp20}).
 \end{example}

In reference~\cite{AU15D}, Alho and Uggla use a dynamical systems analysis to show that global and asymptotic bounds
should be imposed on $-V'(\Phi)/V(\Phi)$ to obtain viable cosmological model that continuously deform $\Lambda$CDM
cosmology. Particularly relevant to the present work are the bounds on that magnitude which
for our positive strictly increasing potentials and our sign convention, translate to $0\leq \lambda<\sqrt{6}$,
or, using~(\ref{lambdaiu}), to $0\leq \mathfrak{v}<1$. Hence, equation~(\ref{eq:lambdaiua}) implies that $\alpha \leq 1$
and the models belong to $\mathcal{C}_{ \alpha}$  if $\alpha <1$. Models with $\lambda \geq \sqrt{6}$, i.e., $\alpha\geq 1$,
(``steep-enough potentials'') do not belong to our class $\mathcal{C}_{\alpha}$ and exhibit an oscillatory behavior
towards the past.~\cite{FO98}
\subsection{Beyond the class $\mathcal{C}_{ \alpha}$: potentials with unbounded separatrices in $\mathfrak{R}_0$}
Although it is true for potentials in the class $\mathcal{C}_{ \alpha}$ (Theorem~2), the notion of separatrix
as given in Definition~1 does not imply boundedness of
$\mathfrak{h}_\mathrm{s}(\varphi) = h_\mathrm{s}(\varphi)/\sqrt{v(\varphi)}$ in $\mathfrak{R}_0$.

For example, let us consider the potential function
\begin{equation}
	\label{emo}
	v(\varphi)= \frac{2\varphi-1}{\varphi^4}\ \, e^{2 \varphi},
\end{equation}
with $\varphi_0>1/2$. It is an exponential function of the type discussed in Proposition~5 modulated
by a decaying rational function. It has been generated by imposing the following solution of~(\ref{hpm1})
\begin{equation}
	\label{modh}
	h(\varphi)=\frac{e^{\varphi}}{\varphi}.
\end{equation}
In this case  $\lim_{\varphi\rightarrow \infty}\mathfrak{v}(\varphi)=1$, so that $v$ is outside the class $\mathcal{C}_{ \alpha}$ with $0\leq \alpha<1$.
Thus, despite the fact that the function $\mathfrak{h}=h/\sqrt{v} \sim \sqrt{\varphi/2}$ is unbounded as $\varphi\rightarrow\infty$,
we will show that it may be considered to be a separatrix.

\begin{figure}
		\includegraphics[width=8cm]{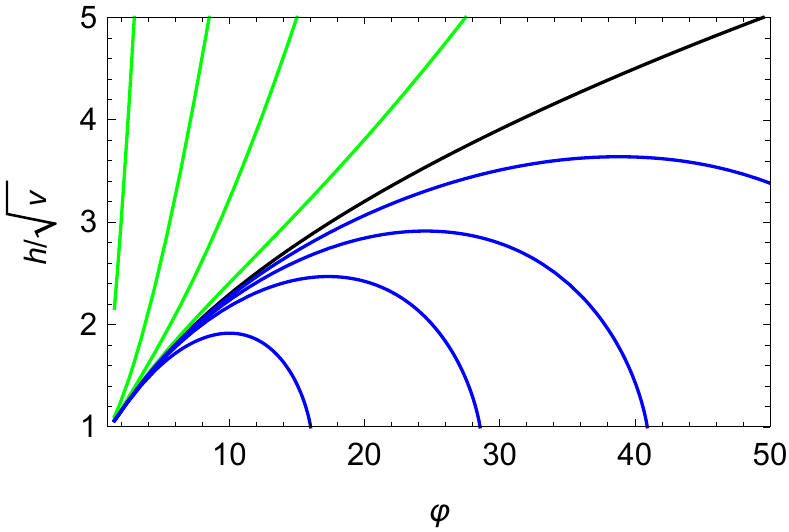}	
	\caption{Unbounded separatrix $\mathfrak{h}_\mathrm{s}(\varphi) = \varphi/\sqrt{2\varphi-1}$ (black curve)
	            and a few trayectories in $\mathfrak{R}_0$ for the potential
	            $v(\varphi)=(2\varphi-1)e^{2\varphi}/\varphi^4$ of equation~(\ref{emo}).\label{fig:fig5}}
\end{figure}

In order to analyze what happens near $h(\varphi)$ we introduce the variable $w\equiv \mathfrak{h}(\varphi)/\sqrt{\varphi}$.
We expect $w$ to be bounded away from zero for the separatrix solution. The equation for $w$ is
\begin{equation}\label{w1}
w'\, \sqrt{\varphi}+ w\frac1{2\sqrt{\varphi}}=\sqrt{\varphi w^2-1}-\mathfrak{v}(\varphi)\sqrt{\varphi}\,w,
\end{equation}
or
\begin{equation}\label{w2}
w'=\sqrt{w^2-\frac1{\varphi}}-\left(1+b(\varphi)\right)w,
\end{equation}
where
\begin{equation}\label{w3}
b(\varphi)\equiv ( \mathfrak{v}(\varphi)+\frac1{2\varphi})-1\sim -\frac1{\varphi}, \quad \varphi\rightarrow \infty.
\end{equation}
Thus, for $\varphi \gg1$ we get
\begin{equation}
	\label{w4}
	2ww'\sim \frac{2w^2-1}{\varphi}.
\end{equation}
An explicit solution is $w_\mathrm{c}\sim 1/\sqrt{2}$ that corresponds to our original function $h$.
The other solutions either go to $+\infty$, if they lie above  $w_c$, or to $-\infty$ infinite
if they lie below  $w_\mathrm{c}$. Indeed, integrating equation~(\ref{w4}) gives
\begin{equation}
	\label{w6}
	|2w^2-1|\sim C\varphi^2.
\end{equation}
For the solutions above (resp.~below) the constant solution $w_\mathrm{c}$ we have $w\sim C\varphi$ (resp.~$w\sim -C\varphi$)
as $\varphi \to\infty$. Hence we have $\mathfrak{h}\sim w\sqrt{\varphi}\sim C\varphi^{3/2}$
so that $h=\mathfrak{h}\sqrt{v}\sim e^{\varphi}$, the maximal  divergence that can be found.
The approximations used in this heuristic argument can be justified because the terms left out in the expansions are of lower order.
We illustrate these results in Fig.~\ref{fig:fig5}, where we plot the unbounded separatrix
\begin{equation}
	\mathfrak{h}_\mathrm{s}(\varphi) = \varphi/\sqrt{2\varphi-1}
\end{equation}
and a few trayectories in $\mathfrak{R}_0$ for the potential~(\ref{emo}).
\section{Asymptotic expansions of separatrices\label{sec:asy}}
 \subsection{Leading term  as $\varphi \rightarrow \infty$: separatrices with backwards inflation}
The next result gives the leading term  of  the asymptotic expansion of the separatrix $h_\mathrm{s}(\varphi)$
as $\varphi \rightarrow \infty$ for potentials in the class $\mathcal{C}_{\alpha}$. In particular, for $\alpha=0$
this leading term  coincides with the slow-roll approximation~\cite{MU05,BA09} to $h_\mathrm{s}$.
\begin{theorem}
The leading asymptotic behavior of the separatrix $h_\mathrm{s}(\varphi)$ for potentials $v\in \mathcal{C}_{\alpha}$ is
\begin{equation}
	\label{asv2}
	h_\mathrm{s}(\varphi)  \sim \frac{\sqrt{v(\varphi)}}{\sqrt{1-\alpha^2}}, \quad \varphi \rightarrow \infty.
\end{equation}
\end{theorem}

\noindent
\emph{Proof:}
For potentials  $v\in \mathcal{C}_{\alpha}$ and $\varphi>\varphi_0$, the function $\mathfrak{h}_\mathrm{s}(\varphi)$ is bounded from below
(by $1$), bounded from above, and satisfies the integral equation~(\ref{inte})
\begin{equation}
	\label{inte2}
	\mathfrak{h}_\mathrm{s}(\varphi)
	=
	\sqrt{\frac{v(\varphi_0)}{v(\varphi)}}
	\mathfrak{h}_\mathrm{s}(\varphi_0)
	\exp{\left(\int_{\varphi_0}^{\varphi}\sqrt{1-\frac{1}{\mathfrak{h}_\mathrm{s}(x)^2}}\, d x\right)}.
\end{equation}
If the potential $v(\varphi)$ diverges as $\varphi \rightarrow \infty$, then from~(\ref{inte2})
it follows that the condition $\mathfrak{h}_\mathrm{s}(\varphi)>1$  can be satisfied  for all $\varphi>\varphi_0$
only if the integral in~(\ref{inte2}) also diverges as $\varphi \rightarrow \infty$.
As a consequence, applying L'H\^opital rule in~(\ref{inte2}) yields
\begin{equation}
	\label{lop}
 	\mathfrak{h}_\mathrm{s}(\varphi)
	\sim
	\frac{\mathfrak{h}_\mathrm{s}(\varphi)}{\mathfrak{v}(\varphi)}
	\sqrt{1-\frac{1}{\mathfrak{h}_\mathrm{s}(\varphi)^2}},
	\quad
	\varphi \rightarrow \infty.
\end{equation}
Hence
\begin{equation}
	\label{int}
	\lim_{\varphi\rightarrow \infty} \mathfrak{h}_\mathrm{s}(\varphi)= \frac{1}{\sqrt{1-\alpha^2}},
\end{equation}
and~(\ref{asv2}) follows.

If the potential $v(\varphi)\rightarrow \mbox{constant}$  as $\varphi \rightarrow \infty$, then $\alpha=0$.
Indeed,  according to Definition~2 $v'(\varphi)>0$ for $\varphi\geq\varphi_0$, and from the elementary equation
\begin{equation}
	v(\varphi) = v(\varphi_0) + \int_{\varphi_0}^\varphi v'(x)\,dx,
\end{equation}
we have that $v'(\varphi)\to 0$ as $\varphi\to\infty$.
Therefore, from~(\ref{inte2}) we deduce  that $\mathfrak{h}_\mathrm{s}(\varphi)$ is bounded only if the integral in~(\ref{inte2})
is convergent as $\varphi \rightarrow \infty$, and this requires that $\mathfrak{h}_\mathrm{s}(\varphi)\rightarrow 1$
($h_\mathrm{s}(\varphi)  \sim \sqrt{ v(\varphi)}\,$)  as $\varphi \rightarrow \infty$. $\Box$

From~(\ref{ih}) and~(\ref{asv2}) it follows  that the separatrices of the models for potentials $v\in \mathcal{C}_{\alpha}$
support inflation as $\varphi\rightarrow \infty$ (backwards  inflation) provided that
 \begin{equation}
 	\label{ih2}
 	\alpha<\frac{1}{\sqrt{3}}.
 \end{equation}
In particular this means that in case the separatrix blows up at a given cosmic time $t^*$,
inflation takes places in a neighborhood of the singularity $t^*$. Again, using~(\ref{eq:lambdaiua}),
condition~(\ref{ih2}) for accelerated expansion is in agreement with the result $\lambda<\sqrt{2}$ of reference~\cite{AU15D}.
\subsection{The asymptotic expansion of the separatrix for divergent potentials with $\alpha=0$}
For potentials such that  $\lim_{\varphi\rightarrow \infty}v(\varphi)= \infty $ and $\alpha=0$
(e.g., monomial potentials), then $v'=o(v)$ and we can go beyond the slow-roll approximation
and find asymptotic expansions of the form
\begin{equation}
	\label{exps}
	h_\mathrm{s}(\varphi) \sim u + \sum_{n=1}^{\infty} \frac{ h_n[u]}{u^n},\quad \varphi \rightarrow \infty,
\end{equation}
where the coefficients $h_n[u]=h_n(u',u'',\ldots,u^{(n)})$ are differential polynomials in the derivatives $u^{(j)}\, (j\geq 1)$ of the function
\begin{equation}
	\label{u}
 	u \equiv \sqrt{v(\varphi)}.
\end{equation}
Indeed, if we substitute~(\ref{exps}) into~(\ref{difin}) and identify coefficients of powers of $1/u^n,\, (n\geq -2)$,
we obtain the recursion relation
\begin{equation}
	\label{recr}
	\sum_{j+k=n; \,j,k\geq -1} \left(\left(h'_j-(j-1) h_{j-1}u'\right)\left(h'_k-(k-1)h_{k-1} u'\right)-h_j\,h_k \right)+\delta_{n,-2}=0,
\end{equation}
where $h'_n$ stands for the total derivative of the differential polynomial $h_n$ with respect to $\varphi$,
and $h_{-2}\equiv 0$. The explicit recursive character of~(\ref{recr}) is exhibited by the equivalent relation
 \begin{equation}
 	\label{recr2}
	h_{n+1}
	=
	\frac{1}{2} \sum_{j+k=n; \,j,k\geq 0} \left( \left(h'_j-(j-1) h_{j-1}u'\right)\left(h'_k-(k-1)h_{k-1} u'\right)-h_j\,h_k \right), \quad n\geq -1.
 \end{equation}
Thus, we find that $h_{-1}=1$, $h_0=0$ and, to third order in $1/u$,
\begin{equation}
	\label{para}
	h_\mathrm{s}(\varphi)
	\sim
	u
	+ \frac{(u')^2}{2u}
	+ \frac{u'' (u')^2}{u^2}
	+ \left(u''' (u')^3+\frac{5}{2}(u''u')^2-\frac{5}{8}(u')^4\right)\frac{1}{u^3}
	+ \cdots,\quad \varphi\rightarrow \infty.
\end{equation}

\begin{example}
For the even monomial potentials
\begin{equation}
	\label{ep}
 	v(\varphi)=\varphi^{2p},
 \end{equation}
equation~(\ref{para}) reduces to a power series expansion
\begin{equation}\
	\label{psp}
	h_\mathrm{s}(\varphi) \sim \varphi^p+\sum_{n=1}^{\infty}b_n\varphi^{p-2n},
\end{equation}
where the coefficients $b_n$ satisfy the recurrence relation
\begin{equation}
	\label{rrm}
	b_1=\frac{p^2}{2},
	\qquad
	b_{n+1} = p(p-2n)b_n-\frac{1}{2}\sum_{j+k=n+1}b_jb_k+\frac{1}{2}\sum_{j+k=n}(p-2j)(p-2k)b_jb_k.
\end{equation}
Thus, the first terms of these expansion are
\begin{equation}
	\label{rrft}
 	h_\mathrm{s}(\varphi)
	\sim
	\varphi^p+ \frac{p^2}{2} \varphi^{p-2}+ \frac{1}{8}p^3(3p-8)\varphi^{p-4}
	+\frac{1}{16}p^4(5p^2-40p+72)\varphi^{p-6}+\cdots.
\end{equation}
In particular, for the quadratic potential ($p=1$) we find,
\begin{equation}
	\label{dc3}
	h_\mathrm{s}(\varphi)
	\sim
	\varphi+ \frac{1}{2\varphi}-\frac{5}{8\varphi^3}+\frac{37}{16\varphi^5}-\frac{1773}{128\varphi^7}+\cdots,
\end{equation}
and for the quartic potential ($p=2$)
\begin{equation}
	h_\mathrm{s}(\varphi) \sim \varphi^2
	                                      + 2
	                                      - \frac{2}{\varphi^2}
	                                      + \frac{12}{\varphi^4}
	                                      - \frac{122}{\varphi^6}+\cdots.
\end{equation}
\end{example}

\begin{example}
For the Higgs potential $v(\varphi)=(\varphi^2-a^2)^2$ the function $u=\varphi^2-a^2$, whose inverse powers
can in turn be re-expanded in powers of $1/\varphi^2$, to give again an asymptotic power series whose first terms are
\begin{equation}
	\label{dc4}
	h_\mathrm{s}(\varphi)
	\sim
	\varphi^2 + 2-a^2 + \frac{2(a^2-1)}{\varphi^2} + \cdots.
\end{equation}
\end{example}
\subsection{Educated match summation of asymptotic expansions in inverse powers of the inflaton}
Similar asymptotic expansions (in particular for the square of the Hubble parameter) have been derived by different procedures,
and there is some interest in the numerical summation of these series, which is typically performed by Pad\'e approximants~\cite{AU15}.
In this brief section we point out how the recently-developed educated match summation method~\cite{AS17} can be used
to advantage for this purpose.

\begin{figure}
		\includegraphics[width=8cm]{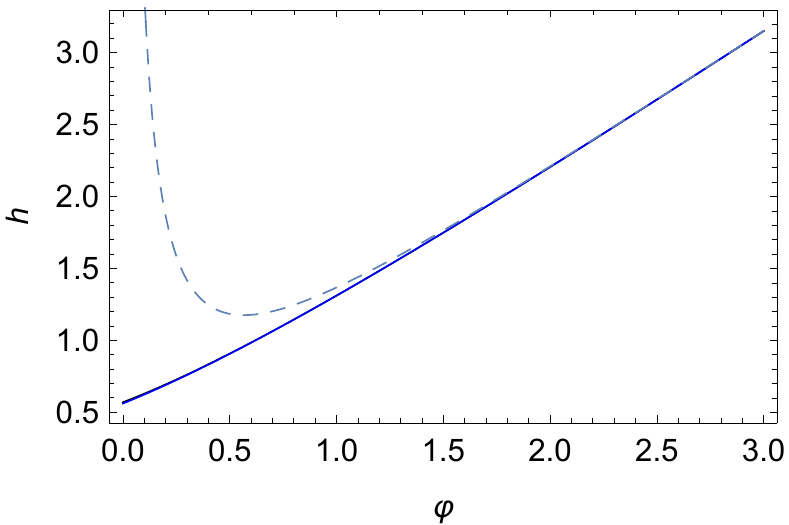}	
	\caption{Separatrix of the quadratic potential
	              $v(\varphi)=\varphi^2$ obtained by numerical integration of the
	              differential equation, by the analytic approximant~(\ref{bor})
	              (continuous lines, actually sumperimposed in the figure),
	              and by a Pad\'e approximant (dashed line).
	              The value at the origin given by the numerical integration
	              is $h_\mathrm{s}(0) = 0.56917264\ldots$, while the
	              value given by the analytic approximant is $h_\mathrm{s}(0)\approx\sqrt{\pi/10} = 0.560499\ldots$,
	              with an error of less than $1.53$\%.\label{fig:fig6}}
\end{figure}

For concreteness, let us consider the asymptotic expansion of the separatrix for the quadratic potential~(\ref{dc3}).
From the recursion relation~(\ref{rrm}) with $p=1$ it follows that
\begin{equation}
	\label{qot}
	\frac{b_n}{b_{n-1}}\sim -2 n,\quad n\rightarrow \infty,
\end{equation}
i.e., in addition to the alternating sign, there is a factorial growth of the coefficients. These two facts lead to conjecture that
the series might be Borel summable, and that the method of educated match, wherein the series to be summed is matched
to the known, Borel-summable asymptotic expansion of (in general, a linear combination of scaled versions of)
the confluent hypergeometric function $\Phi(z)=z^{-a} U(a,1+a-b,1/z)$. We proceed in close analogy to the calculation
of section~3.3 in~\cite{AS17}: because of the pattern of signs in equation~(\ref{dc3}), we pull out a factor $1/\varphi$.
The simplest approximant requires only the first two coefficients, $b_0=1/2$ and $b_1=-5/8$, and we choose
$a=1/2$, $b=1$ to impose regularity of the approximant at $\varphi=0$. With these values of the parameters the
confluent hypergeometric function can be written in terms of the complementary error function $\erfc$
(see~equation~[7.1.2] in~\cite{AS72}) and we find
\begin{equation}
	\label{bor}
	h_\mathrm{s}(\varphi) \sim \varphi+\sqrt{\frac{\pi}{10}} e^{2\varphi^2/5}\erfc(\sqrt{2/5} \varphi).
\end{equation}
This simple, analytic approximation to the separatrix $h_\mathrm{s}(\varphi)$ is surprisingly accurate on all the range $\varphi\geq 0$.
For example, it gives the maximum error at $h_\mathrm{s}(0)\approx\sqrt{\pi/10} = 0.560499\ldots$, while the value obtained by (unstable)
numerical integration is $h_\mathrm{s}(0) = 0.56917264\ldots$, i.e., an error of less than $1.53$\%. The error
decreases monotonically and quickly as $\varphi\to\infty$. As an illustration, in Fig.~\ref{fig:fig6} we plot the
result of a numerical integration of the corresponding differential equation using a shooting strategy to
find the appropriate initial condition at $\varphi=0$, the graph of the approximant~(\ref{bor}), and, in dashed line,
the result of a $[1,1]$ Pad\'e approximant which uses one more term of the expansion~(\ref{dc3}). The first two graphs
are in effect superimposed, while the Pad\'e approximant diverges at the origin. Higher-order approximants are
increasingly accurate.

Likewise, for the quartic potential we find
\begin{equation}
	\label{bor4}
	h_\mathrm{s}(\varphi) \sim 1 + \varphi^2 +\sqrt{\frac{\pi}{2}} \varphi e^{\varphi^2/4}\erfc(\varphi/2).
\end{equation}
This analytic approximation gives the maximum error at $h_\mathrm{s}(0)\approx 1$, while the value obtained by (unstable)
numerical integration is $h_\mathrm{s}(0) = 0.954931\ldots$, i.e., an error of less than $4.72$\%.
\subsection{$\alpha$-attractor E-models and the Starobinsky potential}
Single-field inflationary models with potentials of the form
 \begin{equation}
 	\label{E-model}
	v(\varphi) = (1-e^{-\beta \varphi})^{2p},
	\quad
	(\beta>0),
 \end{equation}
are called $\alpha$-attractor~E-models~\cite{KALI13,FE13,LI13}. In particular the case $p=1$ 
is the well-known Starobinsky model~\cite{ST80,WH84}, which fits nicely the most recent experimental results~\cite{PL18}.

Theorem~2 shows that these models do have separatrices,
but since the function $u=\sqrt{v(\varphi)}$ is bounded as $\varphi \rightarrow \infty$, the ansatz~(\ref{exps}) does not apply.
However, there is a unique asymptotic series of the form
\begin{equation}
	\label{estsb}
	h_{\mathrm{s}}(\varphi) \sim \sum_{n=0}^{\infty}b_ne^{-n\beta\varphi}.
\end{equation}
Indeed, if we substitute this expansion into equation~(\ref{difin}) and identify the coefficients of the exponentials $e^{-n\beta\varphi}$
we get the recurrence relation
\begin{equation}
	\everymath{\displaystyle}
	\sum_{j+k=n}(\beta^2jk-1)b_jb_k=\left\{\begin{array}{ll}
		(-1)^{n-1}\binom{2p}{n} & n=0,1,\dots,2p,\\
		0                       & n>2p.
\end{array}\right.
\end{equation}
The solution to this recurrence relation has to be given independently for three ranges of $n$. Concretely,
the first three coefficients are
\begin{equation}
	\label{rr1}
	b_0=1,\quad b_1=-p,\quad b_2=\frac{p(2p-1)}{2}+\frac{1}{2}(\beta^2-1)p^2;
\end{equation}
for $n=3,\dots,2p$
\begin{equation}
	\label{rr2}
	b_n=-\left(\beta^2(n-1)-1\right)pb_{n-1}+\frac{1}{2}\sum_{j+k=n,j,k\geq2}(\beta^2jk-1)b_jb_k+\frac{(-1)^n}{2}\binom{2p}{n},
\end{equation}
while for $n>2p$
\begin{equation}
	\label{rr3}
	b_n = -\left(\beta^2(n-1)-1\right) pb_{n-1}+\frac{1}{2}\sum_{j+k=n,j,k\geq2}(\beta^2jk-1)b_jb_k.
\end{equation}
Moreover,
\begin{equation}
	\label{qot2}
	\frac{b_n}{b_{n-1}}\sim -\beta^2 n p,\quad n\rightarrow \infty,
\end{equation}
and it is also a reasonable conjecture that the series~(\ref{estsb}) might be Borel-summable in the variable $e^{-\beta\varphi}$.

By way of example, the first three terms of the asymptotic expansion for the separatrix
of the general $\alpha$-attractor~E-model are
\begin{equation}
	\label{esep}
	h_{\mathrm{s}}(\varphi) \sim 1-p e^{-\beta\varphi}+\Big( \frac{p(2p-1)}{2}+\frac{1}{2}(\beta^2-1)p^2\Big) e^{-2\beta\varphi}+\cdots,
\end{equation}
which for the Starobinsky potential~\cite{ST80,WH84} reduce to
\begin{equation}
	\label{tru}
	h_\mathrm{s}(\varphi) \sim 1-e^{-\beta  \varphi}+\frac{\beta^2}{2}\, e^{-2\beta \varphi}+\cdots.
\end{equation}

We may also apply this method for negative values of $\varphi$ by considering the potentials $\tilde{v}(\varphi)=v(-\varphi)$,
which belong to the class $\mathcal{C}_{\alpha}$ with $\alpha=p\beta$ provided that $p\beta<1$.
This condition is equivalent to the condition $\bar{\lambda}<1$ used in reference~\cite{AU17} to study the global dynamics
of E-models.
\subsection{Exponentially steep potential well}
As a final example we briefly study the steep exponential potential
\begin{equation}
	\label{fo}
	v(\varphi) = e^{2\beta\varphi}+e^{-2\beta\varphi},
\end{equation}
where $\beta$ is a positive constant. Foster~\cite{FO98} shows that this potential has a separatrix if $\beta<1$.
This result follows also from our Theorem~2, since this potential belongs to the class $\mathcal{C}_{ \alpha}$ with $\alpha=\beta$.

It is easily seen that the asymptotic expansion for the separatrix can be written as
\begin{equation}
	\label{fohs}
	h_s(\varphi) \sim \frac{e^{\beta\phi}}{\sqrt{1-\beta^2}}+\sum_{n=1}^{\infty}b_ne^{-\beta(4n-1)\varphi}.
\end{equation}
Substitution of~(\ref{fohs}) into~(\ref{hpm1}) leads to the recurrence relation
\begin{equation}
	b_n = \frac{\sqrt{1-\beta^2}}{2\left((4n-1)\beta^2+1\right)}\sum_{j+k=n}\left((4j-1)(4k-1)\beta^2-1\right)b_jb_k,
\end{equation}
and the first terms in the expansion of $h_s(\varphi)$ as $\varphi\rightarrow\infty$ are
\begin{eqnarray}
	\label{sepsteep}
	h_s(\varphi) & \sim &\frac{e^{\beta\varphi}}{\sqrt{1-\beta^2}}+\frac{\sqrt{1-\beta^2}}{2(3\beta^2+1)}e^{-3\beta\varphi}+
	\frac{(1-\beta^2)^{3/2}(9\beta^2-1)}{8(3\beta^2+1)^2(7\beta^2+1)}e^{-7\beta\varphi}\nonumber \\
      & & {}+\frac{(1-\beta^2)^{5/2}(9\beta^2-1)(21\beta^2-1)}{16(3\beta^2+1)^3(7\beta^2+1)(11\beta^2+1)}e^{-11\beta\varphi}+\cdots.
\end{eqnarray}
\subsection{Separatrices with or without blow up inflaton field}
Finally, we will use the asymptotic result~(\ref{para}) to make a brief digression on the behavior  of the inflaton field
$\varphi_\mathrm{s}(t)$ corresponding to a separatrix $h_\mathrm{s}(\varphi)$ as a function of the cosmic time.
These solutions $\varphi_\mathrm{s}(t)$  may be either defined for all $t$ or blow up at a finite negative value of $t$.
For potentials that diverge as $\varphi\rightarrow \infty$ and with $\alpha=0$ we can use the expansion~(\ref{para})
to obtain
\begin{equation}
	\label{est}
	{h}_\mathrm{s}' = \sqrt{h_\mathrm{s}^2-v} \sim (\sqrt{v})',\quad  \varphi \rightarrow \infty.
\end{equation}
Hence, equation~(\ref{impb}) shows that $\varphi_\mathrm{s}(t)$ blows up if and only if
$1/(\sqrt{v})'$ is integrable as $\varphi \rightarrow \infty$.

In particular the separatrices of inflaton models with even monomial potentials $v=\varphi^{2p}$ for $p=1,2$
as well as Higgs potentials $v(\varphi)=(\varphi^2-a^2)^2$ have separatrix filelds $\varphi_\mathrm{s}(t)$
without blow up. For instance, the inflaton or the separatrix of the Higgs model with $a=1$ is given explicitly by
\begin{equation}
	\label{cei2}
	\varphi_\mathrm{s}(t) = \varphi_0 e^{-2t}.
\end{equation}

Likewise, the expansion~(\ref{tru}) of the separatrix for the Starobinski model~(\ref{tru})
shows that $1/\sqrt{h^2-v} \sim e^{\beta \varphi}/\beta$. Therefore the integral~(\ref{impb})
is divergent and the separatrix determines a solution $\varphi_\mathrm{s}(t)$ of~(\ref{eq:me}) without blow-up.

The case $\varphi\to -\infty$ for the Starobinski model can be analyzed with the change of variables~(\ref{cha}),
which reduces the problem to the case $\varphi\to\infty$ for the potential $\tilde{v}(\varphi)=(e^{\beta\varphi}-1)^2$.
It can be proved that a separatrix exists only for $\beta<1$.

Finally, the separatrices $h_\mathrm{s}(\varphi)=e^{\alpha \varphi}$ of to the exponentially increasing
potentials $v_\alpha$~(\ref{exp2})  of class $\mathcal{C}_{ \alpha}$ are associated to the explicit inflaton fields
\begin{equation}
	\label{cei3}
	\varphi_\mathrm{s}(t)=-\frac{1}{\alpha}\log\left(\alpha^2 t+e^{-\alpha \varphi_0 }\right),
\end{equation}
which blow up at a finite time.
\begin{acknowledgments}
The financial support of the Spanish Ministerio de Econom\'{\i}a y Competitividad under Projects No.\ FIS2015-63966-P,
PGC2018-094898-B-I00 and PGC2018-098440-B-I00 is gratefully acknowledged. J.L.V. thanks the Departamento de
An\'alisis Matem\'atico y Matem\'atica Aplicada of the Universidad Complutense de Madrid for his appointment as Honorary Professor.
\end{acknowledgments}
%
\end{document}